# Photoelectrochemical water splitting in separate oxygen and hydrogen cells

Avigail Landman[1], Hen Dotan[2], Gennady E. Shter[3], Michael Wullenkord[4], Anis Houaijia[4], Artjom Maljusch[5], Gideon S. Grader[3] & Avner Rothschild[2]

[1] The Nancy & Stephen Grand Technion Energy Program (GTEP), Technion – Israel Institute of Technology, Technion City, Haifa 32000 Israel.

[2] Department of Materials Science and Engineering, Technion – Israel Institute of Technology, Technion City, Haifa 32000 Israel.

[3] Department of Chemical Engineering, Technion – Israel Institute of Technology, Technion City, Haifa 32000 Israel.

[4] Institute of Solar Research, German Aerospace Center (DLR), Cologne 51147, Germany.

[5] Evonik Creavis GmbH, Marl 45772, Germany.

**Solar water splitting provides a promising path for sustainable hydrogen production and solar energy storage. One of the greatest challenges towards large-scale utilization of this technology is reducing the hydrogen production cost. The conventional electrolyzer architecture, where hydrogen and oxygen are co-produced in the same cell, gives rise to critical challenges in photoelectrochemical (PEC) water splitting cells that directly convert solar energy and water to hydrogen. Here we overcome these challenges by separating the hydrogen and oxygen cells. The ion exchange in our cells is mediated by auxiliary electrodes, and the cells are connected to each other only by metal wires, enabling centralized hydrogen production. We demonstrate hydrogen generation in separate cells with solar-to-hydrogen conversion efficiency of 7.5%, which can readily surpass 10% using standard commercial**

**components. A basic cost comparison shows that our approach is competitive with conventional PEC systems, enabling safe and potentially affordable solar hydrogen production.**

Water electrolysis ($2H_2O \rightarrow 2H_2 + O_2$) combined with renewable power sources such as solar or wind provides a promising path for sustainable hydrogen production for fuel cell electric vehicles (FCEVs)[1] and power-to-gas storage of variable power sources.[2] One of the greatest challenges towards large-scale utilization of these clean energy technologies is reducing the hydrogen production cost. This may be achieved using photoelectrochemical (PEC) water splitting cells that directly convert water and sunlight to hydrogen and oxygen.[3] The current architecture of PEC water splitting cells[4] resembles that of conventional water electrolysis cells,[5] comprising a sealed cell with two electrodes and a membrane that separates the $O_2$ and $H_2$ gas products. This architecture is well-suited for centralized hydrogen production in alkaline or polymer electrolyte membrane (PEM) electrolyzers, but not for distributed solar hydrogen production in PEC cells. The low power density of the sunlight necessitates a large number of PEC cells, giving rise to critical challenges for gas separation, collection and transport to a centralized hydrogen storage and distribution facility safely and affordably. This work addresses these challenges by separating the hydrogen and oxygen production such that the hydrogen can be produced in a central hydrogen generator far away from the solar field where oxygen is produced. The ion exchange between the anode and cathode in the oxygen and hydrogen cells, respectively, is mediates by reversible solid-state redox relays that can be cycled many times with minimal efforts. Our separate cells approach works well for both electrolysis and PEC cells, as demonstrated in the following.

The conventional water electrolysis cell design is illustrated in Fig. 1a, featuring an alkaline electrolyzer.[5] Both electrodes are dipped into one cell containing the aqueous electrolyte, with an ion exchange membrane or porous diaphragm separating the cell into anode and cathode compartments that generate $O_2$ and $H_2$, respectively. The two compartments can be separated using a salt bridge, as illustrated in Fig. 1b, but this is not a common practice because the series resistance of the salt bridge reduces the electrolysis efficiency. Recently, a new PEM electrolyzer architecture was proposed by Rausch *at el.* using a soluble molecular redox mediator (silicotungstic acid) that mediates the electron-coupled proton exchange between the oxygen and hydrogen evolution reactions (OER and HER, respectively).[6] The redox mediator is reduced at the glassy carbon cathode of an electrolytic cell while water is oxidized at the Pt mesh anode, and then transferred to another cell with Pt catalyst that gives rise to spontaneous $H_2$ evolution from the reduced mediator, as illustrated in Fig. 1c. Thus, the OER and HER are decoupled, and $O_2$ and $H_2$ are generated in separate cells. This provides potential advantages, as discussed in ref. 6, but it also has some critical drawbacks and limitations. Firstly, the reported efficiency of the electrochemical process in the electrolytic cell and the Faradaic efficiency for $H_2$ evolution in the second cell were rather low, 63% and 68%, respectively, ending up in an overall efficiency of 43%. This is considerably lower than state-of-the-art PEM electrolyzers that reach up to 68.3% based on the free energy metrics.[5,7] Secondly, the redox mediator is acidic, therefore precious Pt catalysts are required. And thirdly, given the dark color of the redox mediator the feasibility of this approach for PEC solar cells is questionable. To overcome these problems, we sought out a solid-state redox system that could mediate the ion exchange between the anode and cathode in alkaline aqueous solutions and be cycled multiple times. The material selection criteria of the desired redox mediator are discussed in the SI. Accordingly, the NiOOH / Ni(OH)$_2$ redox couple was selected for use as auxiliary electrodes (AEs), as illustrated in Fig. 1d. This idea was first disclosed by us in a patent application filed in 2015.[8] A very recent report by Chen *et al.* demonstrates a similar concept with sequential $H_2$ and $O_2$ generation in two steps,[9] as opposed to continuous co-generation of $H_2$ and $O_2$ in separate cells presented herein.

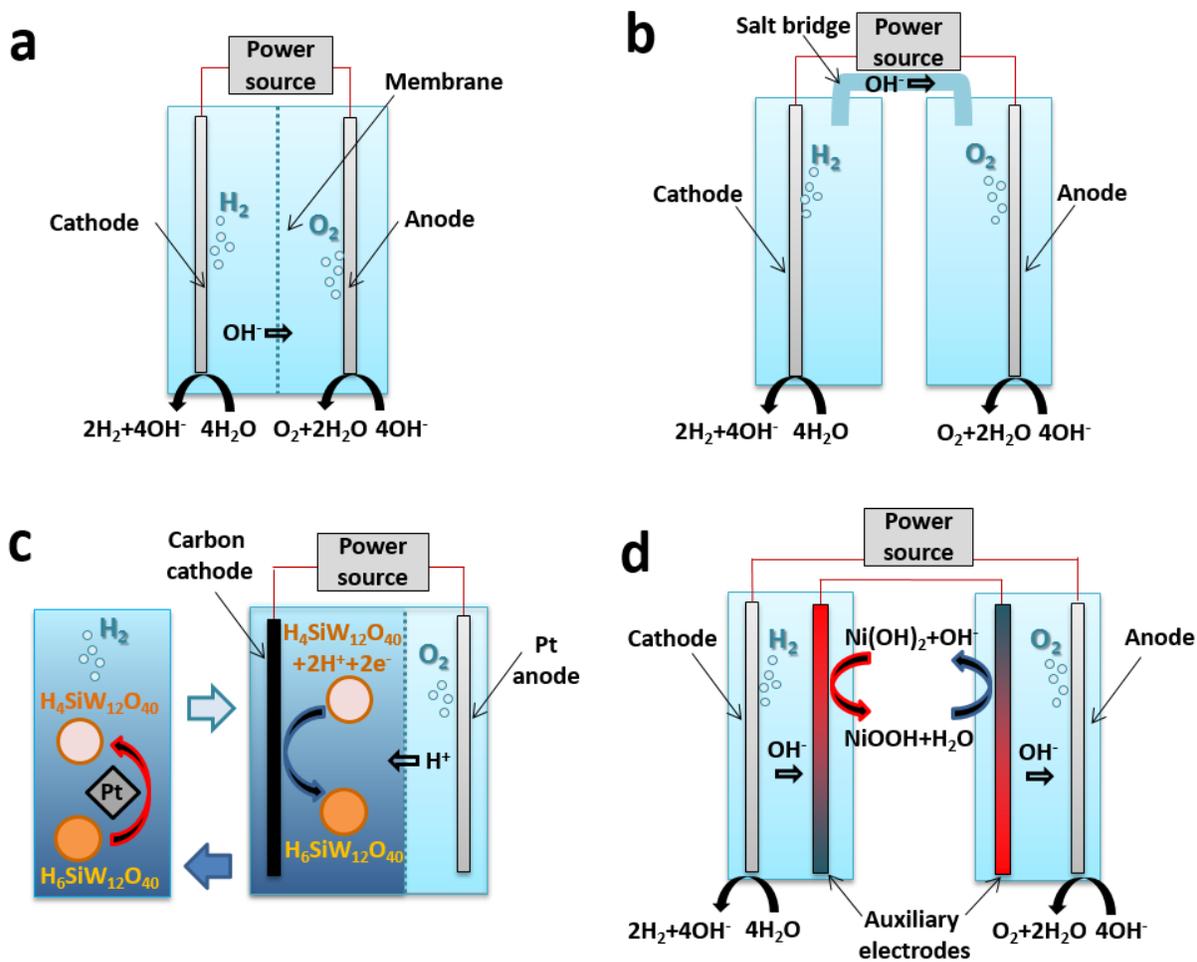

**Figure 1| Water electrolysis cell architectures. a,** Conventional configuration of an alkaline water electrolysis cell. **b,** Same as **(a)** with the membrane replaced by a salt bridge. **c,** New PEM electrolyzer configuration proposed by Rausch *at el*. (*2*) with soluble redox mediator that enables decoupling the $H_2$ and $O_2$ generation steps. **d,** Our membrane-free configuration for alkaline water electrolysis in separate hydrogen and oxygen cells (*3*). The anode can be replaced by a photoanode or a photoanode-PV tandem stack, thus turning the electrolysis cell into a PEC water splitting solar cell that directly convert water and solar power to hydrogen fuel.

NiOOH / Ni(OH)$_2$ electrodes are widely employed in rechargeable alkaline batteries,[10,11] where charged NiOOH electrodes reversibly convert to Ni(OH)$_2$ upon discharging: NiOOH + H$_2$O + e$^-$ ⇌ Ni(OH)$_2$ + OH$^-$.[12] The NiOOH / Ni(OH)$_2$ redox reaction in our AEs precedes the OER by more than 200 mV, as demonstrated in the cyclic voltammograms presented in Fig. S1. This enables the AEs to serve as reversible redox relays that mediate the ion (OH$^-$) exchange with the primary electrodes (anode and cathode) in separate cells, while the OER and HER take place only at the primary electrodes. Thus, O$_2$ and H$_2$ are produced in separate cells with no O$_2$ / H$_2$ crossover. The operation principle of our two-cell water electrolysis system is illustrated in Fig. 1d and in more details in Fig. S2. In the oxygen cell, the OER (4OH$^-$ → O$_2$ + 2H$_2$O + 4e$^-$) occurs at the anode, converting OH$^-$ ions supplied by the AE to O$_2$ gas, while the AE transforms from NiOOH to Ni(OH)$_2$. In the hydrogen cell, the HER (4H$_2$O + 4e$^-$ → 2H$_2$ + 4OH$^-$) occurs at the cathode, and the OH$^-$ ions generated by this reaction are consumed by the AE that transforms from Ni(OH)$_2$ to NiOOH. Thus, one AE charges while the other one discharges. The electrons participating in the OER and HER reactions transfer from the cathode to the anode through a power source that drives the reactions, whereas the electrons participating in the AE redox reactions transfer from one AE to another through another metal wire. Summing all the reactions in the oxygen and hydrogen cells yields the overall water splitting reaction: 2H$_2$O → 2H$_2$ + O$_2$.

Owing to the remarkable cycling durability of the NiOOH / Ni(OH)$_2$ AEs, that can be cycled thousands of times as demonstrated in rechargeable alkaline batteries,[13] electrolysis can be carried out repeatedly by cycling the AEs either by reversing the current polarity or swapping their places. To demonstrate this, Fig. 2 shows 40 electrolysis cycles in separate oxygen and hydrogen cells with Ni foil primary electrodes and commercial Ni(OH)$_2$ AEs. Prior to this test, the AEs were activated through charge-discharge cycles during which the discharge capacity increased from 59% of the transferred charge (22.5 mAh) on the first cycle to 99.6% at the end of the activation process (see Fig. S8). Following the activation process, a charged NiOOH electrode was placed near the anode in the oxygen cell, and a discharged Ni(OH)$_2$ electrode was placed near the cathode in the hydrogen cell. The electrodes were connected by copper wires as illustrated in Fig. 1d, and electrolysis was carried out by forcing a constant current of 45 mA (5 mA/cm$^2$) between the anode and cathode, while measuring the applied voltage, $V_{appl}$. Figure 2a shows $V_{appl}$ as a function of time during 20 h of continuous operation in which the current polarity was reversed whenever the voltage reached a threshold limit of ±3 V. This limit was set to avoid oxygen evolution in the hydrogen cell and vice versa, as explained in the SI.

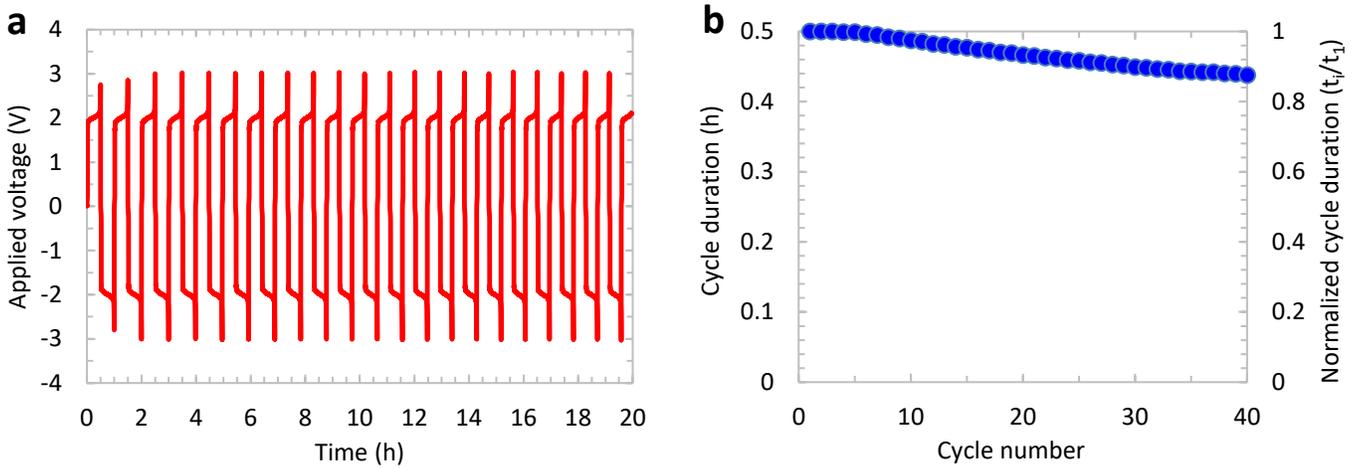

**Figure 2| Two-cell water electrolysis cycles in separate hydrogen and oxygen cells. a,** $V_{appl}$ as a function of time during 20 h of operation at a constant current of 45 mA (5 mA/cm$^2$). **b,** The cycle duration plotted against the cycle number. The measurement was carried out in 1M NaOH alkaline aqueous solution at ambient temperature.

In every cycle, $V_{appl}$ increased gradually from 1.8 V at the beginning of the cycle to 2.3 V close to its end, ending with a sudden jump to the threshold limit (±3 V). The changes in $V_{appl}$ followed the changes in the potential difference between the AEs in the oxygen and hydrogen cells, $\Delta U_{AE} = U_{AE,O_2} - U_{AE,H_2}$, as shown in Fig. S15(b). This resembles the discharge curve of a battery, indicating that the changes in $V_{appl}$ track the state of charge of the AEs. The sudden jumps at the end of the cycles indicate complete discharge of the AEs, which must be recharged to continue the process. Here, recharging is achieved by reversing the current polarity and performing another electrolysis cycle with the oxygen and hydrogen cells being swapped, as illustrated in Fig. S2. In Fig. S12 we demonstrate cycling by swapping the AEs instead of reversing the current polarity as shown here. The two methods are equivalent and yield very similar results.

The first electrolysis cycle in Fig. 2 lasted 30 min before reaching the threshold voltage. Subsequently, the cycle duration decreased by 0.3%, on average, from one cycle to another, reaching 26 min after 40 cycles, as shown in Fig. 2b. This drift is due to incomplete charging of the AE in the hydrogen cell. However, the AE can be recharged back to its initial state whenever necessary as demonstrated in Fig. S13. The cycle duration depends on the initial charge of the AEs and on the electrolysis current. It can be extended by increasing the charge and reducing the current. In Fig. 2 the AEs were charged to 22.5 mAh, a small fraction of their rated capacity (1300 mAh, according to the vendor). Consequently, the cycle duration was short (30 min at 45 mA). In Fig. S10 we show much longer cycles of > 6 h achieved by charging the AEs to 448 mAh. During the whole test (125 h), gas bubbles were formed on the primary electrodes, but not on the AEs (see Video S1). Gas chromatography (GC) measurements confirmed stoichiometric $O_2$ and $H_2$ production in the oxygen and hydrogen cells, respectively, with Faradic efficiency of ~100% and no $O_2$ / $H_2$ crossover up to ~80% of the charged capacity of the AEs (see Figs. S19 and S21, respectively). Thus, the results presented here demonstrate that alkaline water electrolysis with continuous co-production of $H_2$ and $O_2$ gases in separate cells can be carried out using reversible solid-state AEs that can be cycled multiple times.

Next, we examine the electrolysis efficiency of our system and the polarization loss incurred by the AEs. At 100% Faradaic efficiency, the electrolysis efficiency equals to the voltage efficiency, $\eta_V = V_{rev}/V_{appl}$, where $V_{rev} = 1.23$ V is the reversible voltage of water electrolysis (at 25°C). Averaged over a 125 h electrolysis test with 20 cycles of > 6 h (Fig. S10), the applied voltage ($<V_{appl}>$) was 2.12 V, of which 0.12 V was the average voltage drop on the AEs, $<\Delta U_{AE}>$ (see Fig. S15(c)). Thus, the average electrolysis efficiency of our system was 58%. This is an excellent result for a simple electrolysis system with Ni foil electrodes operated under mild conditions (1M NaOH, ambient temperature). The electrolysis efficiency can be readily enhanced by using rare earth OER and HER catalysts such as $RuO_2$ and Pt, respectively, as well as by using concentrated alkaline solution and operating the cell at elevated temperature, as commonly done in alkaline electrolyzers.[5] These routes were not pursued in this study because we aim to demonstrate the effectiveness of our separate cells approach rather than break the efficiency record. Thus, more important than the overall electrolysis efficiency of the entire system is the polarization loss incurred by the AEs. This was, on average, 0.12 V, which is only 5.7% of the applied voltage. This is a reasonable toll, comparable to the Ohmic loss across the membrane in other PEC cell architectures.[14-16] $<\Delta U_{AE}>$ increases with increasing current density, but the increase is slow, and at a current density of 40 mA/cm$^2$ in the hydrogen cell and 4 mA/cm$^2$ in the oxygen cell $<\Delta U_{AE}>$ reaches only 0.16 V, see Fig. S11. It is noteworthy that the polarization loss incurred by the redox reactions of the AEs scales logarithmically with the current, whereas the Ohmic loss across the membrane in conventional electrolysis scales linearly with the current.

Having demonstrated that the hydrogen cell can be separated from the oxygen cell in a simple alkaline water electrolysis system without degrading the efficiency with respect to alternative electrolysis architectures, we show next how this concept can be applied for the design of a hydrogen refueling station with unsealed PEC solar cells or PEC-PV tandem cells[7,17] connected by metal wires to a centralized $H_2$ generator at the refueling station, as illustrated in Fig. 3. The motivation for proposing this disruptive concept becomes clear when considering the requirements to separate and collect the $H_2$ gas produced in a large number of conventional PEC cells distributed in the solar field and transport it to a centralized $H_2$ storage and distribution facility, and the complex measures that must be taken to ensure safe operation in compliance with strict fire protection standards.[18] Considering as a case study a hydrogen refueling station with a production rate of 400 kg $H_2$ / day, an average insolation of 180 W/m$^2$ and a solar to hydrogen (STH) conversion efficiency of 10%, the net area of the PEC solar cells should be > 30,000 m$^2$. Using 10×10 cm$^2$ PEC cells, the largest PEC cells reported to date,[19] means an array of > 3,000,000 cells. To collect the $H_2$ from all these cells they must be sealed and fitted with $H_2$ gas piping manifold. In addition to the immense piping construction, the cells must be fitted with membranes to prevent intermixing of $H_2$ and $O_2$, a highly flammable gas mixture, adding complex engineering and material challenges.[20] Furthermore, in order to comply with the strict fire protection standards of hydrogen technologies,[18] complex safety measures would have to be taken to ensure that no gas leaks occur within the PEC cells and at the joints along the $H_2$ gas piping manifold, and that unexpected accidents are immediately extinguished. The required safety measures give rise to tremendous efforts and expenses, as discussed in the SI. These challenges, in addition to efficiency and stability challenges, render solar $H_2$ production using conventional PEC cells economically questionable.[21,22] Our separate cells architecture overcomes these hurdles in an elegant way that is also very feasible, as demonstrated below.

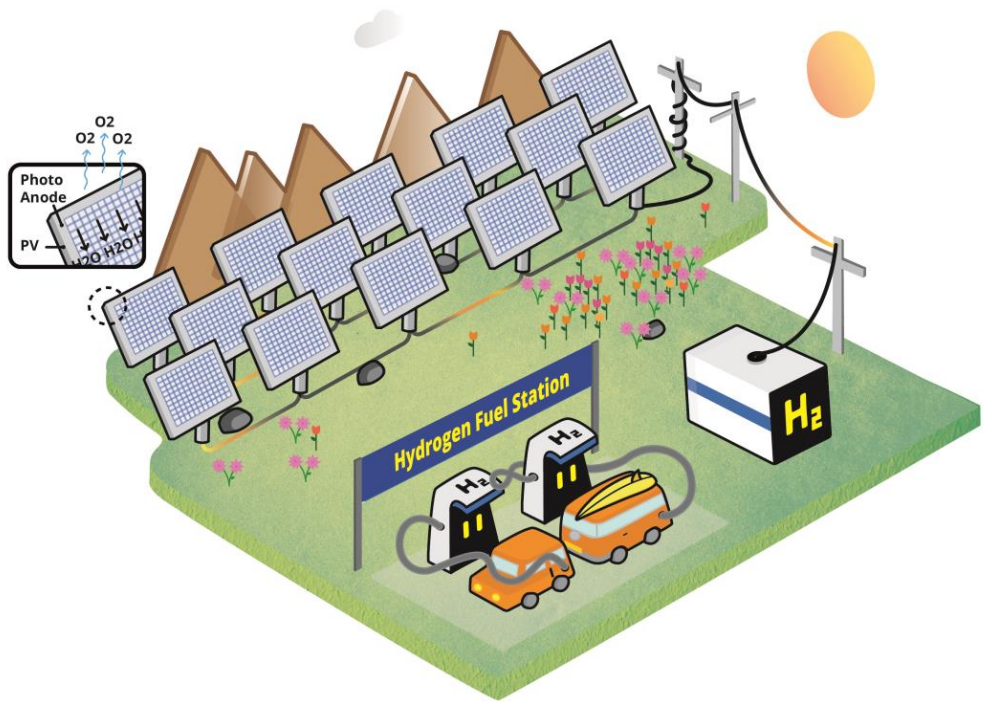

**Figure 3| Conceptual illustration of a solar hydrogen refueling station** with distributed PEC solar cells producing $O_2$ and a centralized $H_2$ generator. A detailed conceptual design of the solar field is illustrated in Fig. S26.

Since the performance of current photoanodes for PEC water splitting is not sufficiently high for their integration with high-end PV cells,[7,23] we demonstrate the utilization of the AEs in a PV-coupled electrolysis system that simulates a PEC-PV tandem cell. The system comprises of a Si PV module connected to oxygen and hydrogen cells with primary Ni foil and Pt-plated stainless steel mesh electrodes, respectively, and activated NiOOH and Ni(OH)$_2$ AEs. The system design follows the constraints that an analogue PEC-PV tandem cell would be subjected to, e.g., the oxygen cell conformally maps the PV module, and the AEs in the oxygen and hydrogen cells are interchangeable. The inset in Fig. 4 shows a schematic illustration of our solar water splitting system. The geometry of the cells is depicted in Fig. S22.

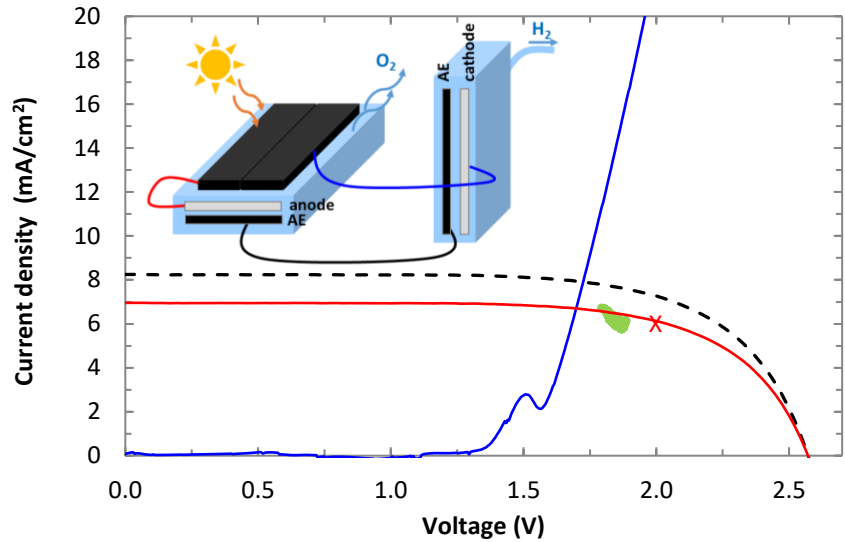

**Figure 4| Solar water splitting system with separate oxygen and hydrogen cells.** Current density – voltage (J-V) characteristics of the individual components, the PV module (red curve) and water electrolysis system (blue curve), and the operation point of the coupled PV-electrolysis system (green dots). The red X marks the maximum power point of the PV module. The black dashed line curve shows the J-V characteristics of a PV module of the same model based on the vendor's specifications. The inset shows schematic illustration of the system.

The current density – voltage (J-V) characteristics of the PV module, measured under simulated solar radiation (AM1.5G), and of the two-cell water electrolysis system are depicted in Fig. 4 by the red and blue curves, respectively. These J-V characteristics were measured separately, prior to connecting the PV module to the electrolysis system. The two curves cross at V = 1.7 V and J = 6.7 mA/cm$^2$. However, the operation point of the coupled system was at slightly higher voltages and lower current densities, as shown by the green dots in Fig. 4. This small deviation is due to additional coupling losses such as wire and contact resistances as well as the voltage drift of the electrolysis system, as was discussed before. It is noteworthy that the short-circuit current density of this particular PV module, $J_{sc}$ = 7.0 mA/cm$^2$, is considerably lower than the vendor's specifications (8.3 mA/cm$^2$).[24] The black dashed line curve in Fig. 4 shows the expected J-V curve of a PV module with $J_{sc}$ = 8.3 mA/cm$^2$.

The STH efficiency of the coupled system is calculated according to the following equation:[23]

$$STH = \left[ \frac{I_{sc}(mA) \times 1.23(V) \times \eta_F}{A_{module}(cm^2) \times P_{in}(mW/cm^2)} \right]_{AM1.5G}$$

where $I_{sc}$ is the short-circuit current of the PV-coupled electrolysis system (measured at standard AM1.5G insolation conditions), $\eta_F$ is the Faradaic efficiency, 1.23 V is the reversible voltage ($V_{rev}$) of the water splitting reaction (at 25°C), $A_{module}$ is the area of the PV module (6.03 cm$^2$), and $P_{in}$ is the power density of the incident light (100 mW/cm$^2$). Taking $\eta_F = 100\%$ (see Fig. S19) and $< I_{sc}/A_{module}> = <J_{sc}> = 6.05$ mA/cm$^2$ (Fig. 4), the STH efficiency was 7.5%, averaged over 1 h of operation. Taking the photoactive area of the PV module (4.9 cm$^2$) instead of the total area (6.03 cm$^2$), which includes inactive area due to interconnects, yields an average STH efficiency of 9.1%. The ratio between the photoactive and inactive area can be readily increased in large-area PV modules, therefore the higher STH value (9.1%) can be readily approached by sizing-up the components. Furthermore, considering the difference in $J_{sc}$ between the particular PV module that was used in this test (7.0 mA/cm$^2$) and the vendor's specifications (8.3 mA/cm$^2$), it is expected that the STH would increase by up to 18.6%, relative to the present result, by selecting a better PV module. This would bring the STH efficiency up to 8.9% based on the total area of the PV module, or 10.8% based on the net photoactive area. Further improvement would be possible by tailoring the J-V characteristics of the PV module to cross the J-V characteristics of the electrolysis system at its maximum power point, e.g., using a DC/DC power converter. Assuming a 90% DC/DC power conversion efficiency, this would bring the STH efficiency up to 11.7%, based on the net photoactive area (see Fig. S24).

Despite the use of a PV module of rather low-performance that does not optimally match the electrolysis system, and primary electrodes that have not been optimized for maximum electrolysis performance, the efficiency of our system is comparable to state-of-the-art solar water splitting systems including both buried junction[25,26] and PV-coupled electrolysis configurations,[27-29] see Table 1. Although higher STH efficiencies have been reported, they were obtained using high-efficiency non-commercial[28] or expensive double-junction[25,26] or even triple-junction concentrator PV cells,[29] rare-earth catalysts,[26,29] and without product separation.[26-28] More important than the STH efficiency, which depends greatly on the PV cell used to drive the water splitting reaction, is the normalized STH / PCE ratio that is independent of the PV efficiency.[7] Based on this figure of merit, our system scores respectively high (see 7$^{th}$ column in Table 1). Besides efficiency, another important feature of solar water splitting systems that has great impact on their safety and cost is whether the hydrogen production is distributed or centralized. Examples of distributed systems are buried junction[25,26] and PEC-PV tandem cells;[7,17] whereas centralized systems are present in PV-coupled electrolysis[27-29] and our separate cells architecture. Centralized hydrogen production has important operational advantages in terms of $H_2$ gas collection, process compactness, safety, etc., whereas PEC-PV tandem cells offer potential advantages in efficiency[7] and cost.[21,30] Our separate cells approach yields the best of both worlds: it enables centralized hydrogen production, as illustrated in Fig. 3, yet it also works for PEC-PV tandem cells, as demonstrated in the following.

**Table 1. Comparison between our results and state-of-the-art solar water splitting systems that have been reported recently.**

| Device configuration | PV cell | PCE (PV) | OER / HER Catalysts | \<STH\> / time | \<STH\> / PCE(PV) | Product separation | Hydrogen production | Reference |
|---|---|---|---|---|---|---|---|---|
| Buried junction | GaInP / GaAs | 13.9% | Ni / NiMo | 8.6% / 2 h | 61.9% | Yes | Distributed | 25 |
| Buried junction | GaInP / GaInAs | 27% | $RuO_2$ / Rh | 12.3% / 16 h | 45.6% | No | Distributed | 26 |
| PV-coupled electrolysis | Si | 16% | $NiB_i$ / NiMoZn | 9.7% / 7 d | 60.6% | No | Centralized | 27 |
| PV-coupled electrolysis | $CH_3NH_3PbI_3$ | 17.3% | $NiFe(OH)_2$ / $NiFe(OH)_2$ | 12.3% / 75 s | 71.1% | No | Centralized | 28 |
| PV-coupled electrolysis | GaInP / GaAs / GaInAsSb | 39.2% | Ir / Pt | 30% / 48 h | 76.5% | Yes | Centralized | 29 |
| PV-coupled electrolysis | Si | 12.3% | Ni / Pt | 7.5% / 1 h | 61.0% | Yes | Centralized | This work |

Figure 4 demonstrates a standalone solar water splitting system with separate hydrogen and oxygen cells driven by a PV module. Similarly, it is possible to construct a PEC-PV tandem system by using PEC oxygen cells with semiconductor photoanodes instead of metal anodes, as illustrated in Fig. S26. We have not done this because our photoanodes are not sufficiently good for coupling with high-end PV cells. While this challenge is being addressed elsewhere,[31-33] we wanted to show here that a two-cell PEC water splitting system could work just as well as a conventional PEC cell. To this end we constructed a two-cell system with a thin film hematite ($\alpha$-$Fe_2O_3$) photoanode[34] in the oxygen cell, a Pt foil cathode in the hydrogen cell, and NiOOH / $Ni(OH)_2$ AEs. The J-V characteristics of this two-cell PEC configuration were nearly the same as a single cell configuration using the same photoanode and cathode without a separator (Fig. S25). This result demonstrates that the separation of the oxygen and hydrogen cells can be achieved in a PEC cell just as well as in an electrolysis cell.

The separation of $H_2$ and $O_2$ production into two cells with only electrical connections between them offers important advantages over other PEC cell architectures. First and foremost, the $H_2$ is produced in a centralized $H_2$ generator that can be placed at the end-user's location, far away from the solar field, as illustrated in Fig. 3. This eliminates the need to seal the PEC cells and fit them with complicated and costly $H_2$ gas piping manifold, thereby simplifying the construction of solar water splitting plants employing a large number of PEC cells. Secondly, the separation of the hydrogen and oxygen cells enables safe operation with no $H_2$ / $O_2$ crossover, without membranes. Otherwise, complex measures would have to be taken to ensure safe operation, including not only the membrane separators and gas-tight sealing of each and every one of the cells, but also continuous pressure monitoring of the entire $H_2$ gas piping manifold, periodic leak tests, continuous $H_2$ gas detection at critical joints, and other safety measures in compliance with the strict fire protection standards of hydrogen technologies.[18] In terms of efficiency, the overpotential toll for driving the reversible redox reaction of the AEs is rather small, ~150 mV (see Fig. S11). This toll amounts to 7.5 – 10% of the total bias voltage that drives the water splitting reaction (1.5 – 2 V). On the bright side, the AEs substitute the membrane separator in the conventional PEC cell architecture which gives rise to an Ohmic overpotential of ~50-100 mV,[14] and reduces the incident light intensity by >10%. In addition, our open cell architecture solves the problem of gas bubble crowding at the front window that reduces the light intensity at the photoelectrode. Therefore, the efficiency of our PEC cell architecture is in par with or even higher than that of the conventional PEC cell architecture.

In terms of cost, it is difficult to make a precise techno-economic analysis and estimate the levelized hydrogen production cost because there are too many unknowns and uncertainties. A rigorous techno-economic analysis builds upon a detailed plan of the solar water splitting plant, which is currently lacking. Such an analysis is beyond the scope of this work, and it shall be carried out elsewhere. However, a basic comparison of the costs saved by substituting the membrane separators and $H_2$ gas piping manifold by the AEs, including their daily swapping during the lifetime of the plant, shows the potential of our approach to offer tremendous savings in the overall balance, as discussed in the SI. Likewise, our approach offers potential economic and safety advantages for long-distance transport of solid-state AEs vs. $H_2$ gas from the solar plant to the end user, see SI. Therefore, our membrane-free water splitting approach in separate oxygen and hydrogen cells, as presented and demonstrated in this work, brings PEC water splitting closer to industrial application. Future efforts should focus on optimizing the AE composition, cell architecture and cycling process in order to improve the performance and reduce the capital and operating costs of PEC solar water splitting plants.

**Methods**

Our two-cell alkaline water electrolysis systems with separate oxygen and hydrogen cells were constructed according to the illustration in Fig. 1d, with primary electrodes (i.e. cathode and anode) connected to a power source and battery-grade $Ni(OH)_2$ / $NiOOH$ auxiliary electrodes (AEs, purchased from Batterix, Israel) connected to each other by a copper wire. A charged NiOOH AE was placed in the oxygen cell and a discharge $Ni(OH)_2$ AE was placed in the hydrogen cell. The electrolyte was 1 M NaOH in deionized water for all the electrolysis tests, and all the tests were carried out in ambient temperature. The AEs were activated before each test by constant current charge-discharge cycles until their capacity stabilized. Cyclic voltammetry measurements of a fully charged NiOOH electrode were carried out in a three-electrode configuration with an Hg/HgO/1M NaOH reference electrode at different scan rates (1, 5, 20, 50 and 100 mV/s), see Fig. S1. For the low-current density test presented in Fig. 2, the primary electrodes were made of nickel foil. Electrolysis cycles were carried out by applying a constant current between the primary electrodes while measuring the applied voltage ($V_{appl}$), until reaching the threshold voltage limit of 3 V. Once the threshold voltage was reached, the current polarity was reversed to continue electrolysis. In parallel to measuring $V_{appls}$, the potentials of the AEs were also measured with respect to Hg/HgO/1M NaOH reference electrodes in their respective cells by two additional electrometers (see Fig. S14). In an alternative cycling method, the electrolysis cycles were carried out by swapping the AEs when the threshold voltage limit was reached (see Fig. S12). For the high-current density test presented in Fig. S11, an iridium-plated titanium plate and a platinum-plated titanium mesh were used as the anode and cathode, respectively. The system was operated by applying a constant current of 128 mA but the different electrodes had different areas such that the current densities at the anode, cathode, discharging AE and charging AE were 8.13, 80, 40 and 4 mA/cm$^2$, respectively. For the PV-electrolysis test presented in Fig. 4, a nickel foil and platinum-coated stainless steel mesh were used as the anode and cathode, respectively. The power source was a PV module comprising four Si PV cells connected in series (IXOLAR solar module SLMD121H04, IXYS, Republic of Korea) that was placed on top of the oxygen cell and illuminated by solar simulated radiation (1000 W/m$^2$). The only power source was the PV module, with no additional external bias. Finally, for the photoelectrolysis test presented in Fig. S25, a hematite (α-Fe$_2$O$_3$) photoanode and Pt sheet cathode were used. The hematite photoanode was prepared as described elsewhere.[34] A linear sweep voltage scan was performed between 0.5 – 2 V in the dark, under solar simulated illumination, and under chopped-light conditions. The measurements were conducted once in a single-cell configuration with no separator, and once in a two-cell configuration using Ni(OH)$_2$ / NiOOH AEs. The hydrogen and oxygen purity in the respective cells was measured by gas chromatography analysis for gas samples which were extracted from the cells at constant time intervals. The Faradaic efficiency was evaluated for both cells by measuring the volume of the effluent gases using a total flow meter. A detailed description of the experimental methods is presented in the SI.

**Supplementary Information** is available in the online version of the paper.

**Acknowledgments** The research leading to these results has received funding from the Solar Fuels I-CORE program of the Planning and Budgeting Committee and the Israel Science Foundation (Grant No. 152/11), from the Israeli Ministry of National Infrastructure, Energy and Water Resources, and from Europe's Fuel Cell and Hydrogen Joint Undertaking (FCH-JU) under Grant Agreement n. [621252]. The results were obtained using central facilities at the Technion's Hydrogen Technologies Research Laboratory (HTRL), supported by the Nancy and Stephen Grand Technion Energy Program (GTEP) and the Adelis Foundation. A. R. acknowledges support for developing photoelectrodes and PEC-PV tandem cells for solar water splitting from the European Research Council under the European Union's Seventh Framework Programme (FP/2007-2013) / ERC Grant Agreement n. [617516]. G. E. S. acknowledges support from the Committee for Planning and Budgeting of the Council for Higher Education under the framework of the KAMEA Program. G. S. G. acknowledges support from the Arturo Gruenbaum Chair in Material Engineering. The authors thank Yeshayahu Lifshitz and Scott C. Warren for reading the manuscript and providing useful suggestions for improving it.

**Author Contributions** A.R. and G.G. conceived and guided the entire project. A.L., H.D. and G.S. designed the experiments. A.L. and H.D. performed the experiments and analyzed the data. All authors contributed to the cost analysis. A.R., G.G. and A.L. wrote the manuscript. All authors discussed the results and commented on the manuscript.

**Author Information** Reprints and permissions information is available at www.nature.com/reprints. The authors declare no competing financial interests. Readers are welcome to comment on the online version of the paper. Correspondence and requests for materials should be addressed to A.R. (avner@mt.technion.ac.il) and G.G. (grader@technion.ac.il).